# *Rāga* Identification using Repetitive Note Patterns from prescriptive notations of Carnatic Music

Ranjani H. G. and Thippur V. Sreenivas
Dept of ECE, Indian Institute of Science, Bangalore-12, India
ranjani@iisc.ac.in, tvsree@iisc.ac.in

**Abstract**

Carnatic music, a form of Indian Art Music, has relied on an oral tradition for transferring knowledge across several generations. Over the last two hundred years, the use of prescriptive notations has been adopted for learning, sight-playing and sight-singing. Prescriptive notations offer generic guidelines for a *rāga* rendition and do not include information about the ornamentations or the *gamakas*, which are considered to be critical for characterizing a *rāga*. In this paper, we show that prescriptive notations contain *rāga* attributes and can reliably identify a *rāga* of Carnatic music from its octave-folded prescriptive notations. We restrict the notations to 7 notes and suppress the finer note position information. A dictionary based approach captures the statistics of repetitive note patterns within a *rāga* notation. The proposed stochastic models of repetitive note patterns (or SMRNP in short) obtained from *rāga* notations of known compositions, outperforms the state of the art melody based *rāga* identification technique on an equivalent melodic data corresponding to the same compositions. This in turn shows that for Carnatic music, the note transitions and movements have a greater role in defining the *rāga* structure than the exact note positions.

## 1. Introduction

Carnatic music is a form of Indian Art music which has been traditionally passed down through generations. It has largely followed an aural tradition. Early major works of 9th century such as *Matanga Muni's Bhrihaddesi* contain solfège notation, akin to western music (Doraiswamy, 2014). However, this notation has not been widely adopted. In the 17th century, Carnatic music became codified the way we know it today through the work *Chaturdandi Prakashika* by *Venkatamakhin* (Sriram, 2004). It is in this work, the melodic framework referred to as *rāgas*, has been mathematically defined. While *rāga* renditions are associated with florid dynamism across notes, the notations used are prescriptive in nature. These notations form the basic skeletal structure of a composition and are usually regarded as generic guidelines for a learner [1]. Thus, the renditions of compositions are replete with *gamakas* (which involve pitch modulations, glides and other ornamentations) while the notations of compositions offer just basic skeletal structure. Some related socio-political and musicological history is chronicled in (Weidman, 2006), (Sambamoorthy, 1998).

The role of prescriptive notations in conveying the essence of a *rāga* within a composition has been debated. Many Carnatic musicians are of the view that Carnatic music cannot be notated as it follows an aural tradition, and even if notated would not serve much purpose (Vijayakrishnan, 2007) [2]. In his detailed musicological study, Vijayakrishnan (2007) presents analogy between Carnatic music and grammar. The work expresses that notation is not indicative of many nuances of *rāgas*, but motivates towards need for a detailed (descriptive) notation. Contrarily, in (Subramanian, 2012), it is argued that the skeletal notation allows a performer to interpret the composition according to their own music tradition. This can result in multiple ways of rendering a compositional notation. There have been inceptive semi-automated sys-tems such as the 2 component stage-dance synthesis model (Subramanian et al., 2012) and the "Gayaka" system (Subramanian, 2012) to generate acceptable rendition(s) given the notations. Thus, the significance of skeletal structure of prescriptive notations in defining *rāgas* in Carnatic music has been irresolute and inconclusive. This paper addresses a scientific curiosity - 'Do prescriptive notations contain any information about *rāga* structures?'[3].

---

[1] Descriptive notations (followed in western classical music) offer a precise rule, wherein the rendition exactly follows the note sequences in the notation.

[2] cf. http://www.kgvijayakrishnan.com/notation.html

We present a brief overview of notations used in Carnatic music in Section 1.1 before motivating the reader towards Carnatic music analysis from skeletal notations (Section 1.2).

### 1.1 An Overview of Carnatic Music Notation

Similar to the notes in solfège notation of western music, the notes in Carnatic music are referred to as the *swaras*. They are represented as $\{Sa, Ri_1, Ri_2, Ri_3, Ga_1, Ga_2, Ga_3, Ma_1, Ma_2, Pa, Da_1, Da_2, Da_3, Ni_1, Ni_2, Ni_3\}$. They correspond to 12 pitch positions in an octave with $Sa$ (referred to as *shadja*) representing the tonic. The relative frequency ratios of these notes are presented in Appendix A. A *rāga* can use atleast 4 notes from the set $\{Sa, Ri, Ga, Ma, Pa, Da, Ni\}$ with, $Ri \in (Ri_1, Ri_2, Ri_3)$, $Ga \in (Ga_1, Ga_2, Ga_3)$, $Ma \in (Ma_1, Ma_2)$, $Da \in (Da_1, Da_2, Da_3)$ and $N \in (Ni_1, Ni_2, Ni_3)\}$ (Rāman, 2008). Table 2 shows the notes within an octave used in some *rāgas*.

*Rāgas* contain structures at 3 layers - (i) note positions (ii) *gamakas* (iii) note movement. A *rāga* rendition involves usage of selected notes along with those *gamakas*, which are characteristic transitions between notes in a melody. The skeletal notations of a composition contain only information pertaining to note positions and note movement.

A typical notation for a particular composition in *rāga Sriranjani* is shown in Figure 1. Table 1 details the current notations used in Carnatic music [4]. A typical notation contains a sequence of notes and the associated lyrics. It can also be viewed as a sequence of notes (or lyrics) set within multiple metric cycles (*āvarthanas*). We emphasize that Carnatic music notation comprises of note sequences with coarse durational representation, while *gamakas* are not represented (Subramanian, 2013). Most notations contain a couple of commencement lines indicating the finer note positions involved in the chosen *rāga*. However, the finer note positions are not indicated for the rest of the composition. Such notations have been successfully used for learning, sight-singing and sight-playing.

### 1.2 Motivation towards analysis of notations

Computational analysis of *rāgas* involves identification, synthesis or motif spotting for *rāgas*. Existing

---

BROCHEVAREVARE
-Thyagaraja

| $Ma,$ | $MaGaRi,$ | $Ri\,Ga$ | $GaRi$ | $Ri,$ | $Sa;$ | $Ni$ | $Sa$ | $Ri$ | $Ga,$ | |
|---|---|---|---|---|---|---|---|---|---|---|
| bro | che | vaa | re | va | re | ra | ghu | pa | the | |

| $\underline{Ma};\,Da$ | $MaGaRi,$ | $RiGaMaGa$ | $GaRi$ | $Ri,$ | $Sa;$ | $Ri$ | $Ga$ | $Ma$ | $Da$ | $Ni$ |
|---|---|---|---|---|---|---|---|---|---|---|
| bro | che | vaa | re | va | re | ni | nu | vi | naa- | |

| $Ni\,Da\,Ma,$ | $MaGaRi,$ | $RiGaMaGa$ | $GaRi$ | $Ri,$ | $Sa;$ | $Ri$ | $Ga$ | $Ma$ | $\underline{Da}$ | $\dot{Sa}$ | $Ni$ | $Da$ |
|---|---|---|---|---|---|---|---|---|---|---|---|---|
| bro | che | vaa | re | va | re | sri | ra | ma | ne | na | ru | na |

**Figure 1:** [Color online] An illustration of musical notation of a composition in *Sriranjani rāga* composed by Thyagaraja (Source: http://ananthp.github.io/carnatic_scores/). Three metric cycles of the transcription are shown. The text in black denotes the notes, and those in blue denotes the composition's lyrics. We highlight that the note sequence does not include finer pitch position information.

| Indian music notations | Explanation |
|---|---|
| *Swaras Sa, Ri, Ga,* | Whole-notes similar to |
| *Ma, Pa, Da, Ni* | solfège notation in Western music |
| ‖ | End of a metric cycle |
| \| | A sub-measure of a metric cycle |
| *tāla* | Musical meter |
| *āvarthana* | Repeated cycle of *tāla* |
| $\dot{X}$ or $\underset{\cdot}{X}$ | A note $X$ in higher or lower octave |
| $\underline{X}$ or $\underline{\underline{X}}$ | The note $X$ is a semi-note or quarter-note |
| $X,\ X;\ X,;\ \&\ X;;,$ | Increase duration of note $X$ by one, two, three, & six beats |

**Table 1:** Carnatic music notation alongside its explanation

approaches for computational analysis of *rāgas* has mainly involved music recordings (Gulati et al., 2014, 2015; Dutta and Murthy, 2014; Gulati et al., 2016; Dutta et al., 2015). The large abundance of audio recordings aid towards such approaches. However, there has been little work to analyze *rāgas* from their notations (Ross et al., 2017). This is more pronounced in Carnatic music due to usage of *gamakas* in renditions of compositions but seldom find representations in notations. It is a popular belief that *gamakas* can be used for distinguishing *rāgas* (Vijayakrishnan, 2007; Krishna and Ishwar, 2012) [5]. A study and analysis of *gamakas* shows that they contain short-term structural information (Krishnaswamy, 2003; Krishna and Ishwar, 2012). We hypothesize longer structures of *rāgas* are reflected in the note movements which are present in prescriptive notations. Albeit musical scores/notations could be associated with *rāga* information, there are pedagogical, music exercises recommended in acknowledged music references such as (Sambamoorthy, 1998) which chal-lenge learners/enthusiasts to identify a *rāga* from only 7 notes and prescriptive note movements; this is a non-trivial task (such as identifying topic/author from text) and suggests that it necessitates good knowledge of the semantics of a *rāga*. While grammar has been strongly associated with Carnatic music, there is possibilities of presence of partial grammatical structures even within the prescriptive notations. With most of the compositions being associated with prescriptive notations, we present a computational analysis that is possible with such (prescriptive) information; the approach as well as data is complementary to that in melodic contours. In this paper, we propose approaches to analyze *rāgas* from structures present in notations (which are devoid of *gamakas*) thus addressing a gap in the literature. Further, this analysis presents an alternative schema for similarity analysis of *rāga* structures using notational/score based information.

---

[3]Throughout this work, we use the term structure to refer to arrangement of notes within a *rāga*.

[4]We use these notations consistently throughout this work. The Indian and Western terminologies here are analogous and not exact counterparts.

[5]for example, *rāga Thōdi* and *Sindhu Bhairavi* are said to have similar note sequences, but are distinguished by *gamakas*

|    | *Rāga* (Abbreviations) | *Swaras* |
|----|------------------------|----------|
| 1. | Hari-Kāmbhōji (Hk)     | $Sa, Ri_2, Ga_3, Ma_1, Pa, Da_2, Ni_2$ |
| 2. | Bhairavi (Bh)          | $Sa, Ri_2, Ga_2, Ma_1, Pa, Da_1, Da_2, Ni_3$ |
| 3. | Shankarābharana (Sb)   | $Sa, Ri_2, Ga_3, Ma_1, Pa, Da_2, Ni_3$ |
| 4. | Thōdi (Td)             | $Sa, Ri_1, Ga_2, Ma_1, Pa, Da_1, Ni_2$ |
| 5. | Nātta (Na)             | $Sa, Ri_3, Ga_3, Ma_1, Pa, Da_3, Ni_3$ |
| 6. | Panthuvarāli (Pv)      | $Sa, Ri_1, Ga_3, Ma_2, Pa, Da_1, Ni_3$ |
| 7. | Madhyamāvathi (Md)     | $Sa, Ri_2, Ma_1, Pa, Ni_2$ |
| 8. | Khamas (Kh)            | $Sa, Ri_2, Ga_3, Ma_1, Pa, Da_2, Ni_2$ |
| 9. | Bēgada (Bg)            | $Sa, Ri_2, Ga_3, Ma_1, Pa, Da_2, Ni_2, Ni_3$ |
| 10.| Kalyani (Ka)           | $Sa, Ri_2, Ga_3, Ma_2, Pa, Da_2, Ni_3$ |
| 11.| Sahanā (Sh)            | $Sa, Ri_2, Ga_3, Ma_1, Pa, Da_2, Ni_2$ |
| 12.| Reetigowla (Rg)        | $Sa, Ri_2, Ga_2, Ma_2, Pa, Da_2, Ni_2$ |

**Table 2:** Various *rāgas* and their set of notes within an octave. These do not convey any sequencing information. Entries 3 and 9 have same set of notes, but are melodically different.

### 1.3 This paper: *Rāga* Identification based on Sub-sequence Statistics

In order to analyse *rāgas* and capture structures in notations, we resort to chunking theory (Gobet et al., 2001; Janata and Grafton, 2003; Rohrmeier and Rebuschat, 2012). A chunk is often defined as a collection of elements having strong associations with each other, but weaker associations with elements within other chunks (Gobet et al., 2001). Chunks can be of fixed size or variable sizes. Notes in music can be seen to be analogous to words in natural language processing. In the latter, words combine (grammatically) to form meaningful phrases, sentences to convey an idea or topic; while in music, note sequences form musicological phrases to impart a *rāga bhāva* (emotion) to the listener, and thus a *rāga* can be compared to a language (Vijayakrishnan, 2007). Thus, analogous to computing the word statistics to characterize a language (Biemann and Mehler, 2014), we compute the statistics of sub-sequences of notes to characterize the structure of *rāgas*. The focus of this work is to (i) stochastically model (fixed length as well as variable length) recurrent note patterns (RNP) from notations using 7 basic notes ($Sa, Ri, Ga, Ma, Pa, Da, Ni$) (ii) quantitatively measure agreements of these patterns with musicologically valid reference phrases (MVRP) (iii) test the impact of RNP as a useful representation for *rāga* identification (RID) task (iv) compare the performance of state-of-the-art melody-based RID algorithm on an analogous audio rendition dataset.

The paper is organized as follows: The mathematical formulation required for computing the statistics of sub-sequences in a *rāga* is presented in Section 2. The stochastic models used for computing the statistics of sub-sequences are described in Section 3. Using the sub-sequence statistics we propose a dictionary based approach for *rāga* identification in Section 4. The experimental setup with details on the database and the performance measures are detailed in Section 5. The performance analysis of the proposed *rāga* identification methods based on notations and their comparison with audio based *rāga* identification is presented in Section 6. The relation between the *rāga* identification algorithms is presented in Section 7. The discussion on the results and the algorithms considered are presented in Section 8.

### 1.4 Proximity to symbolic processing in Western and folk music

There have been computational approaches proposed to discover repeated music patterns from symbolic music representations. These have been used to find motifs, themes, chorus and other musical structures on data ranging from Jazz music to Western art music; an overview of the approaches, challenges in attempting musical pattern discovery, using either approximate or exact pattern matches, and in evaluating them in contrast to an expert musician's view point has been addressed in detail in (Janssen et al., 2013). Geometric approaches involving using notion of sequences as a shape have been found to be advantageous for polyphonic music (Meredith et al., 2002). String based approaches (involving string of notes/ pitch class/ MIDI note numbers or XML data) generally involve finding P length pattern in a melody M by sliding all possible P patterns over M and tracking the matches and filtering it using either rules motivated by Generative Theory of Tonal Music theory (Nieto and Farbood, 2012) or its frequency of occurrance/length/duration/rhythmic consistency (Meek and Birmingham, 2001). Other approaches involve finding long repeating patterns (Cambouropoulos, 2006), partitioning methods (Cambouropoulos, 2006),(Karydis et al., 2007). We observe that most approaches involve trees to represent the patterns/ phrases along with the associated frequency of occurrence information (Janata and Grafton, 2003). While many of these approaches use frequency of occurrence to discover patterns, we need to compare the frequency of patterns across *rāgas* in order to compare and contrast the structures present. Hence, we choose to use stochastic models.

## 2. Mathematical formulation

Let the set of all possible notes for any *rāga* be denoted by the basic 7 notes, $\mathbb{V} = \{Sa, Ri, Ga, Ma, Pa, Da, Ni\}$. Let a notation of a *rāga* composition, $r$ be **A**. It comprises of a sequence of notes within metric cycles such that $\mathbf{A} = [A_1, A_2, A_3, \ldots, A_I]$, where $A_i$ is a sequence of notes in $i^{th}$ metric cycle denoted by $A_i \triangleq [u_{t=1}^{(i)}, u_{t=2}^{(i)}, \ldots, u_t^{(i)}, \ldots, u_{t=T_i}^{(i)}] \ \forall i \in \{1, 2, \ldots, I\}$ and $I$ is the number of metric cycles[6]. $u_t^{(i)}$ is the $t^{th}$ note in the $i^{th}$ metric cycle. $u_t^{(i)} \in \mathbb{V}$ and $T_i$ is the number of notes in $A_i$ i.e., $T_i = |A_i|$ where $|.|$ denotes cardinality of a set. The vocabulary $\mathbb{V}$ is the same for all *rāgas*. Since we are interested in computing the frequency of repetition of sub-sequences of notes for characterizing a *rāga*, we refer to any note sub-sequence as "Repetitive Note Patterns" or RNP in short. We propose to discover RNP and their statistics from **A** using stochastic models. Therefore we define the probability of a *rāga* transcript as $\Pr(\mathbf{A}) = \Pr([A_1, A_2, \ldots, A_I])$. Assuming statistical independence across metric cycles [7], we have,

$$\Pr(\mathbf{A}) = \prod_{i=1}^{I} \Pr(A_i) = \prod_{i=1}^{I} \Pr(u_1^{(i)}, u_2^{(i)}, \ldots, u_{T_i}^{(i)}) \quad (1)$$

A sub-sequence **a** of a sequence **A** is denoted by $\mathbf{a} \preceq \mathbf{A}$.

The data models for estimating RNP from **A** are presented in the following section.

## 3. Discovering RNP from music notation

### 3.1 $N$-gram model

A typical stochastic model used in music for modeling fixed length subsequences is the $N$-gram and its varied smoothing techniques (Downie, 1999; Scholz et al., 2009; Unal et al., 2012; Hillewaere et al., 2012; Gillick et al., 2010; Sentürk, 2011). Consider the classical $N$-gram model as applied to note transcripts of a *rāga*, **A**. A graphical model depicting the Markov dependency for tri-gram ($N = 3$) is shown in Figure 2. The $N$-gram model presumes the statistical dependency of a given note on the $N - 1$ previous notes. This will lead to modeling of the dependency of note sequences up to a fixed length of $N$. The probability of **A** as for this model (from (1)):

$$\Pr(\mathbf{A}) = \prod_{i=1}^{I} \Pr(u_1^{(i)}, \ldots, u_N^{(i)}) \prod_{t=N+1}^{T_i} \Pr(u_t^{(i)} | u_{t-N+1}^{(i)} \ldots u_{t-1}^{(i)}) \quad (2)$$

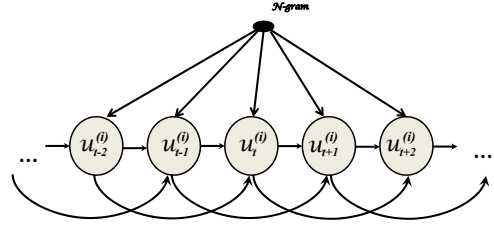

**Figure 2:** A graphical model depicting tri-gram language model.

The first term of (2) is the joint probability of an $N$-length sequence while the second term is the conditional probability of a note, given its $N - 1$ predecessors. When $T_A < N$, the joint probability is approximated using the marginal probabilities.

Consider the set of all possible $N$-length sub-sequences, $\mathbf{S} = \{\mathbf{s}_m\}$ where $\mathbf{s}_m = [y_1^{(m)}, y_2^{(m)}, \ldots, y_N^{(m)}] \in \mathbb{V}^N$, $y_n^{(m)} \in \mathbb{V}$, $m \in [1, 2, \ldots, |\mathbb{V}^N|]$ and $\mathbb{V}^N$ is the $N$-fold Cartesian product. Let $\theta_m \triangleq \Pr(\mathbf{s}_m)$ be the joint probability of the sequences and $\zeta_m \triangleq \Pr(y_N^{(m)} | [y_1^{(m)}, \ldots, y_{N-1}^{(m)}])$ be the conditional probability of the sequence. $\theta_m$ can be estimated from **A** normalized counts of $\mathbf{s}_m$ in **A**. $\zeta_m$ can be found by normalizing across sequences having the same context $[y_1^{(m)}, \ldots, y_{N-1}^{(m)}]$.

We refer to collection of $\{\mathbf{s}_m, \theta_m, \zeta_m\}, \forall m \in \{1, 2, \ldots, |\mathbb{V}^N|\}$ as the $N$-gram dictionary, $\mathcal{D}_{N-gram}$. Now, **S** represents the RNP with their probability of occurrence denoted by $\boldsymbol{\theta}_{N-gram} \triangleq \{\theta_m\}_{m=1}^{|\mathbb{V}^N|}$ and $\boldsymbol{\zeta}_{N-gram} \triangleq \{\zeta_m\}_{m=1}^{|\mathbb{V}^N|}$. To account for unseen sequences, Good-Turing smoothing technique (Gale, 1995) is used. This results in assigning non-zero (and miniscule) probabilities for unencountered sub-sequences.

### 3.2 $M$-skip-$N$-gram model

Analogous to dropping adjectives and descriptives in a sentence without changing the meaning of a sentence, in music also one can skip over or add a few notes without significant perceptual difference in its melodic effect. For example, in a composition, *Swaminatha* of *rāga Nāṭai*, the notes - $Sa, SaRiGaMa$ can be improvised as $SaNiSaRiGaMa$. A statistical model to account for such skips in temporal data is the "$M$-skip-$N$-gram model" (Goodman, 2001),(Guthrie et al., 2006).

We propose an alternative formulation of $M$-skip-$N$-gram model to that in (Guthrie et al., 2006), using the concept of partially ordered sets (posets) (Rudin, 1991). Consider the binary relation, $\leq$, on a set of temporal indices of $A_i$, $\mathbf{t}^{(i)} = \{1, 2, \ldots, T_i\}$. $(\mathbf{t}^{(i)}, \leq)$ is a poset. Let $\tau_k^{(i)}$ be any $N$-element subset of $\mathbf{t}^{(i)}$, defined as $\tau_k^{(i)} \triangleq \{t_{k_1}^{(i)}, t_{k_2}^{(i)}, \ldots, t_{k_N}^{(i)}\}$, with $t_{k_j}^{(i)} \in \mathbf{t}^{(i)}, \forall j \in \{1, 2, \ldots, N\}$, such that $\left(\tau_k^{(i)}, \leq\right)$ is also a poset (Rudin, 1991). Hence, $k \in \{1, 2, \ldots, \binom{T_i}{N}\}$. Let $\boldsymbol{\tau}^{(i)}$ be a collec-

---

[6] Superscripts are within parenthesis to differentiate from exponent operator.

[7] The statistical independence across metric cycles results in chunking synchronized to metric cycles. It offers computational advantages and incorporates the possible boundaries of musical patterns to coincide with (either beginning or end of) the metric cycle. However, not all phrase patterns need be bound to metric cycles.

tion of $N$-element posets such that:

$$\boldsymbol{\tau}^{(i)} = \left\{ \tau_k^{(i)} \mid \left( t_{k_N}^{(i)} - t_{k_1}^{(i)} \right) \leq (N + M - 1) \right\}$$

Given **A**, we define collection of temporal indices, $\mathcal{T}_{M,N}$, for $M$-skip-$N$-gram model as $\mathcal{T}_{M,N} \triangleq \cup_{i=1}^{I} \left\{ \boldsymbol{\tau}^{(i)} \right\}$. Let **U** be the set of $N$-length RNP obtained from collecting the notes in **A** at the temporal indices, $\mathcal{T}_{M,N}$. The cardinality of $\mathcal{T}_{M,N}$ is dependent on $T_i$, $I$, $M$ and $N$. (Guthrie et al., 2006) discusses in detail the coverage possible with $M$-skip-$N$-gram. $N$-gram is a special case of $M$-skip-$N$-gram with $M = 0$.

Let $M$-skip-$N$-gram dictionary, $\mathcal{D}_{(M,N)-skip}$ be a collection of $\{\mathbf{s}_m, \theta_m, \zeta_m\}, \forall m \in \{1, 2, \ldots, |\mathbb{V}^N|\}$. The joint probability $\theta_m$ is estimated from counts of the sequence $\mathbf{s}_m$ in **U** and conditional probabilities are obtained by suitably normalizing probabilities.

There still can exist unseen sequences which are assigned a small non-zero probabilities through Good Turing smoothing. Again, the dictionary $\mathcal{D}_{(M,N)-skip}$ contains $|\mathbb{V}^N|$ number of $N$-length RNP and their associated joint and conditional probabilities.

### 3.3 $N$-multigram model

The use of $N$-gram or $M$-skip-$N$-gram results in fixed length RNP. Since we do not know apriori the length of RNP that can characterize a *rāga*, we would like to explore the flexibility of discovering variable length RNP. $N$-gram based approaches estimate the dependency of a note only based on notes preceding it and does not consider notes succeeding it. However, we want to consider the entire notation into account to discover varied length RNP. We assume that varied length RNP are the building blocks of a notation. This leads to segmentation approach for discovering variable length RNP. A model with similar assumption as addressed in language and text modeling in (Deligne and Bimbot, 1995; Jian-Tao Sun, 2006) is the multigram model. Adopting a similar approach, we segment **A** to $\Omega$ number of consecutive varied length segments, $\{\boldsymbol{x}_j\}_{j=1}^{\Omega}$, such that maximum length of any segment is $N$ i.e., $|\boldsymbol{x}_j| \leq N$. The notations are naturally segmented at the level of *āvarthana* (similar to the sentence boundary information in language models). Including this information, we can now define $\Omega = \sum_{i=1}^{I} \omega_i$, where $\omega_i$ is the number of variable length segments of $A_i$.

Let $\boldsymbol{B}_i$ denote the set of all possible partitions of $A^{(i)}$ resulting in sub-sequences with a maximum length $N$. Thus, there can be a total of $\sum_{k=\kappa}^{T^{(i)}-1} \binom{T^{(i)}-1}{k}$ partition possibilities where $\kappa = \lfloor T^{(i)}/N \rfloor$. A specific partition $\boldsymbol{b}_i = \left[b_0 = 0, b_1, \ldots, b_{\omega(\boldsymbol{b}_i)} = T^{(i)}\right] \in \boldsymbol{B}_i$ results in $\omega(\boldsymbol{b}_i)$ sub-sequences. Let $k^{th}$ sub-sequence be $\boldsymbol{x}_k^{(i)} = [u_{b_{k-1}+1}, \ldots, u_{b_k}]$. Thus, $A^{(i)} = \left[\boldsymbol{x}_1^{(i)}, \boldsymbol{x}_2^{(i)}, \ldots, \boldsymbol{x}_{\omega(\boldsymbol{b}_i)}^{(i)}\right]$. Assuming that $\{\boldsymbol{x}_k^{(i)}\}$ are independent, we have:

$$\Pr(\mathbf{A}|\mathfrak{B} = \mathfrak{b}) = \prod_{i=1}^{I} \prod_{k=1}^{\omega(\boldsymbol{b}_i)} \Pr(\boldsymbol{x}_k^{(i)}) \quad (3)$$

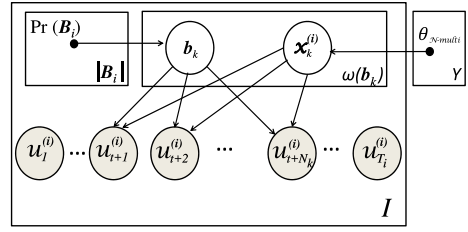

**Figure 3:** A graphical model of $N$-multigram model for a *rāga* $r$.

where $\mathfrak{B}$ denotes the set of all possible boundaries of **A**, and $\mathfrak{b} = \bigcup_{i=1}^{I} \boldsymbol{b}_i$ is one such partition. Such a partition $\mathfrak{b}$, includes the metric cycle boundaries. We refer to this as $N$-multigram model, where $N$ is indicative of maximum length of any $\boldsymbol{x}_k^{(i)}$. Now, the problem is to obtain the boundaries $\mathfrak{b}$, given **A**. This can be solved using explicit search of all possible partitions. The sentence boundary information is discarded in (Deligne and Bimbot, 1995). In contrast, in this work, by including the metric cycle boundaries within $\mathfrak{b}$, the boundary search space is significantly reduced. As in (Deligne and Bimbot, 1995), we employ segmental-K-means to obtain optimal partition $\mathfrak{b}^{(*)}$.

In order to define objective function in segmental-K-means framework, we require the notion of an $N$-multigram dictionary for a *rāga* $r$, $\mathcal{D}_{N-multi}$ which contains the set of all possible note sequences with a maximum length of $N$ and their associated probabilities i.e.,

$$\mathcal{D}_{N-multi} = \left\{ (\boldsymbol{s}_m, \theta_m) : \boldsymbol{s}_m \in \bigcup_{n=1}^{N} \mathbb{V}^n; \theta_m \triangleq \Pr(\boldsymbol{s}_m) \right\}_{m=1}^{Y}$$

where $Y = \sum_{n=1}^{N} |\mathbb{V}|^n$. Defining, $\Theta \triangleq \{\theta_m\}_{m=1}^{Y}$, we can now rewrite (3) which is parameterized by $\Theta$ as:

$$\Pr(\mathbf{A}|\mathfrak{B} = \mathfrak{b}; \Theta) = \prod_{m=1}^{Y} \prod_{i=1}^{I} \prod_{k=1}^{\omega(\boldsymbol{b}_i)} [\theta_m]^{\delta(\boldsymbol{x}_k^{(i)}, \boldsymbol{s}_m)} \quad (4)$$

where $\delta(\boldsymbol{x}_k^{(i)}, \boldsymbol{s}_m) = 1$ if $\boldsymbol{x}_k^{(i)} = \boldsymbol{s}_m$ and 0 otherwise. A graphical model of the same is shown in Figure 3.

With $\Pr(\mathfrak{B})$ assumed to be uniform, the objective function of maximizing complete likelihood, $\Pr(\mathfrak{B}, \mathbf{A}; \Theta)$ in segmental-K-means framework is equivalent to:

$$\max_{\mathfrak{b}, \Theta} \log \Pr(\mathbf{A}|\mathfrak{B} = \mathfrak{b}; \Theta)$$

subject to $\sum_{m=1}^{Y} \theta_m = 1$, so that $\Theta$ is a valid distribution. The unknowns to be estimated are $\mathfrak{b}$ and $\Theta$. The optimal $\mathfrak{b}$ and $\Theta$ can be estimated by alternative maximization hill-climbing approach (Van Trees, 2002). The optimal segmentation given the parameters is estimated, followed by optimal parameter esti-mation for the given segmentation, iteratively as:

$$\mathfrak{b}^{(l+1)} = \bigcup_{i=1}^{I} \underset{\boldsymbol{b}_i \in \boldsymbol{B}_i}{\arg\max} \ \log \Pr(A_i|\boldsymbol{b}_i; \Theta^{(l)}) \quad (5)$$

$$\begin{aligned}
\Theta^{(l+1)} &= \underset{\Theta}{\arg\max} \left\{ \log \Pr(\mathbf{A}|\mathfrak{b}^{(l+1)}; \Theta) \right\} \\
&= \frac{c_m(\mathfrak{b}^{(l+1)})}{\Omega(\mathfrak{b}^{(l+1)})}
\end{aligned} \quad (6)$$

where $l$ indicates iteration index, $c_m(\mathfrak{b}^{(l+1)})$ and $\Omega(\mathfrak{b}^{(l+1)})$ indicates number of times sequence $s_m$ repeats and total number of segments obtained, respectively, given $\mathfrak{b}^{(l+1)}$. $\Theta$ obtained from (6) is a multinomial distribution.

Suppose $s_m = [\mathbf{a} \ b]$ and its sub-sequence $s_k = [\mathbf{a}]$ have same occurrence counts. This implies $\theta_m$ and $\theta_k$ will be same. The subsequent segmentation step will result in a boundary favoring $s_m$ over $s_k$ in order to maximize (5). Thus, the formulation favors segmentation having longer recurrent sub-sequences.

The algorithm iterates between (5) and (6). The stopping criteria is chosen as the rate of change in likelihood between successive iterations below a threshold, $\epsilon$. Since RNP not in $\mathbf{A}$ will necessarily be assigned zero probability and $\sum_{m=1}^{Y^{(r)}} \theta_m = 1$ constraint is valid for every iteration, initial $\mathcal{D}^{(r)}_{N-multi}$ can contain all $\leq N$-length sub-sequences of $\mathbf{A}$. After convergence, RNP of $\mathcal{D}^{(r)}_{N-multi}$ which are associated with zero probabilities are discarded. If $Y_{eff}$ is the total number of RNP associated with non-zero probabilities, post-convergence, $Y^{(r)}_{eff} < \sum_{n=1}^{N} |\mathbb{V}^n|$. No smoothing technique is utilized. The role of dictionaries in RID task is explored in Section 4.

## 4. RNP based *Rāga* Identification

In this section, we propose a method to utilize the discovered RNP and their associated probabilities for RID. Let $\mathcal{R}$ be a set of *rāgas* considered. For each *rāga* $r \in \mathcal{R}$, we obtain a dictionary, $\mathcal{D}^{(r)}$ using the techniques detailed in Section 3 on training data of each *rāga*, where $\mathcal{D}^{(r)} \in \left\{ \mathcal{D}^{(r)}_{N-gram}, \mathcal{D}^{(r)}_{(M,N)-skip}, \mathcal{D}^{(r)}_{N-multi} \right\}$.

Let $\mathbf{A}_{test} = \left[ A_1^{(test)}, A_2^{(test)}, \ldots, A_J^{(test)} \right]$ denote the notation of an unknown *rāga*, with $J$ metric cycles. We propose closed-set *rāga* identification, for a given $\mathbf{A}_{test}$, in ML framework using $\mathcal{D}^{(r)}$ as:

$$\hat{r} = \underset{r \in \mathcal{R}}{\arg\max} \Pr(\mathbf{A}_{test} \ ; \ \mathcal{D}^{(r)}) \quad (7)$$

where $\hat{r}$ is the detected *rāga* identity. We collectively refer to the three stochastic models based approaches for discovering RNP as "SMRNP". The details of estimating $\Pr(\mathbf{A}_{test} \ ; \ \mathcal{D}^{(r)})$ for each of the 3 techniques, are as below.

### 4.1 $N$-gram and $M$-skip-$N$-gram models

Let $\mathcal{D}^{(r)}_{N-gram}$ be the dictionary of a *rāga* $r$ obtained using $N$-gram approach which contains $N$-length RNP and their associated joint ($\boldsymbol{\theta}^{(r)}_{N-gram}$) and conditional probabilities ($\boldsymbol{\zeta}^{(r)}_{N-gram}$). The probability of $\mathbf{A}_{test}$, given $\mathcal{D}^{(r)}_{N-gram}$, can be expanded using chain rule as in (2) for $J$ metric cycles. While discovering RNP from training data, the parameters $\boldsymbol{\theta}^{(r)}_{N-gram}$ and $\boldsymbol{\zeta}^{(r)}_{N-gram}$ are estimated from the data, whereas for the RID task, these parameters are used from $\mathcal{D}^{(r)}_{N-gram}$ for computing the objective function in (7). For $M$-skip-$N$-gram model, $\Pr(\mathbf{A}_{test}; \mathcal{D}^{(r)}_{(M,N)-skip})$ can be estimated similarly.

### 4.2 $N$-multigram model

Given $\mathbf{A}_{test}$, we hypothesize that segmenting it using RNP from $\mathcal{D}^{(r)}_{N-multi}$ would be optimal in an ML sense, if *rāga* corresponding to $\mathbf{A}_{test}$ is indeed $r$. Therefore, we obtain optimally segment $\mathbf{A}_{test}$, w.r.t each $r$ as :

$$\begin{aligned}
\mathfrak{b}^{(*,r)} &= \underset{\mathfrak{b}}{\arg\max} \Pr(\mathbf{A}_{test}|\mathfrak{b}; \Theta^{(r)}_{N-multi}) \\
&= \bigcup_{j=1}^{J} \underset{\boldsymbol{b}_j \in \boldsymbol{B}_j}{\arg\max} \ \log \Pr(A_j^{(test)}|\boldsymbol{b}_j; \Theta^{(r)}_{N-multi})
\end{aligned}$$

While discovering RNP using $N$-multigram, we use alternate maximization method for optimal segmentation and optimal parameter estimation; whereas, for RID, optimal segmentation of $\mathbf{A}_{test}$ using $\Theta^{(r)}$ alone is done. $\Pr(\mathbf{A}_{test}|\mathfrak{b}^{(*,r)}; \mathcal{D}^{(r)}_{N-multi})$ can be calculated using (4) for $J$ metric cycles. The superscript $r$ in $\mathfrak{b}^{(*,r)}$ indicates *rāga* index and not iteration index as in (5).

## 5. Experimental Setup

### 5.1 Carnatic music notation database

There are very few choices for large corpora of Carnatic music notation that are publicly available. One such choice as noted in (Srinivasamurthy et al., 2014) is the "Carnatic Music Archive", which is a steadily growing database compiled by Dr. Shivkumar Kalyanaraman, a practicing Carnatic musician (Shivkumar, 1998). It contains notations along with audio recordings. We choose 12 diverse *rāgas* as listed in Table (II) in this work. The chosen *rāgas* cover almost all note positions and are diverse in the use of *gamakas* during rendition. The chosen *rāgas* are also popular. The number of compositions for each of the chosen *rāgas* varies from 4 to 10. The chosen compositions range from basic *rāga* exercises such as *geethe* to significant ones such as *varna* and *krithi*; each of these can comprise anywhere between 10 to 100 *āvarthanas* of skeletal notes. The dataset is pre-processed before training/testing as detailed below.

### 5.2 Pre-processing

In this work, all the (seven) notes are constrained to a single octave by mapping upper as well as lower octave to middle octave. To investigate the role of note sequences alone for *rāga* identification, we assume each note to be of unit duration. Elongated notes are repeated by the number of durational units; for example, a notation such as "$Ma;$" is replaced with $Ma\ Ma\ Ma$ (cf. Table 1). Some metrical cycles comprise of just $||;||$ - these imply an elongated note present in the previous metric cycle is continued; these are discarded, to incorporate metric cycle independence assumption (cf. 1).

### 5.3 Training and testing setup

We use 4-fold cross-validation setup; the set of compositions of a *rāga* are partitioned into 4 sub-samples. The experiment is repeated 4-folds, with each sub-sample as test set, while the remaining 3 form the training set. The compositions corresponding to a training fold are concatenated resulting in $\mathbf{A}_{train}^{(r)}$. A training fold of each *rāga* comprises of ~200 metric cycles, and consisting of more than 2000 notes. Test set comprises of 30 to 100 metric cycles and 300 to 1500 notes. The partitions are maintained across algorithms to ensure similar training and testing data. The performance of the algorithms is evaluated on each of the 4-folds of the data and the average is reported.

### 5.4 Algorithms considered for comparison of RID

Alongside the performance of SMRNP algorithms for RID, we also investigate role of musicologically accepted phrases as characterizing a *rāga* for RID. We refer to the characteristic phrases of each *rāga* as agreed upon by three trained musicians as the "musicologically valid reference phrases" or MVRP in short. We perform RID on the equivalent melody dataset (which contains the *gamakas* and fine duration information) using the state of art TDMS (Time Delay Melodic Surface) technique proposed in (Gulati et al., 2016). The implementation details of the proposed SMRNP algorithms, MVRP on notation database and TDMS are given in the following sections.

*5.4.1 Details of SMRNP algorithms*

Since each fold of $\mathbf{A}_{train}^{(r)}$ results in a dictionary representation using SMRNP algorithms, each *rāga* has 4 dictionary representations.

The performance is evaluated for $N$-gram model for varied $N$ (up to $N = 8$); it is chosen to represent length of typical MVRP. Since there is no prior work on Carnatic music transcripts, we consider performance of bi-gram, $N = 2$ as baseline performance.

For $M$-skip-$N$-gram model, the number of possible $N$-length sub-sequences considered increases with *āvarthana* length $T$ and with increasing skips, $M$. However, increasing $M$ is seen to skew the distribution towards unlikely sequences, while decreasing importance of adjacent notes. Hence, we report performance for $M = 1, 2$ for bigram and trigram.

For $N$-multigram model, the stopping criteria in the training stage is chosen as rate of change in likelihood to be less than $0.1\%$. A flat initialization of uniform probability distribution is used as $\Theta^{init,r}$ for obtaining $\mathcal{D}_{N-multi}^{(r)}$. In testing stage, for any $\mathbf{A}_{test}$ sequence of $T_{test}$ elements from $\mathbb{V}$ (finite and closed set), there exists atleast one ♭ which partitions $\mathbf{A}_{test}$ into $T_{test}$ subsequences of length one; thus unseen data is handled without smoothing. The performance is reported for $N$ varied from $5$ to $8$.

*5.4.2 Overview of TDMS algorithm and audio database*

The performance of melody-based TDMS for RID algorithm is evaluated alongside the notation-based RID algorithms, for the considered list of *rāgas* and its compositions. In this a 2-dimensional feature is obtained from melody contours as follows:

- Consider melodic sequence $c_t$, where $c_t$ is tonic normalized melody contour (in cent scale) and $t \in \{1, 2, \ldots, T\}$.
- An octave wrapping quantizer with $\eta$ number of levels is constructed; resolution of each level is $1200/\eta$. Mathematically, $v_t = B(c_t) = \left\lfloor \left(\frac{\eta c_t}{1200} \mod \eta\right) \right\rfloor$
- A surface $S$ is created such that each element of the surface $s_{i,j}$ is populated using the quantized melody as, $s_{i,j} = \sum_{t=\tau}^{T} I(v_t, i) I(v_{t-\tau}, j)$ where $0 < i, j < \eta$, $I$ is an indicator operator.
- The surface post-processed : compressed, smoothed across bins and normalized.

The original work reports *rāga* recognition accuracy of 86.7% using TDMS technique on a dataset of 480 full-length high quality recordings of Carnatic music spread over 40 *rāgas* with pitch extracted using Essentia library (https://github.com/MTG/essentia). The parameters used in the original paper are: time delay of $\tau = 0.3s$ to generate the time-delayed melody surface, compression parameter $\alpha = 0.75$, Gaussian smoothing parameter $\sigma = 2$, symmetric KL Divergence as distance measure and $k = 1$ for 1-NN classifier.

In this work, we use an audio database obtained from (Shivkumar, 1998) and corresponding to the notations used for evaluating SMRNP algorithms - we refer to it as equivalent audio database. Pitch is extracted using Praat software (Boersma and Weenink, 2016; Meer and Rao, 2006). The tonic for each composition is manually determined. All melody contours are tonic normalized to $146.83\ Hz$. The same 4-fold training/testing partitions as used with the notation dataset is used. The parameters of TDMS technique in this work are: $\tau = 0.3\ s$, with symmetric KL Divergence as the distance measure, $\sigma = 2$, $\alpha = 0.8$ and 1-Nearest Neighbour classifier ($k = 1$).

*5.4.3 MVRP for Carnatic music*

The set of MVRP of a *rāga* is consistent across musicians but not closed (Srikantan, 2004), (Vasanthamadhavi, 2005). For each of the 12 *rāgas* considered, the MVRP as agreed upon three performing musicians is tabulated and can be accessed at https://sites.google.com/site/carnaticmvrp/ .The musicians were asked to list what they consider as im-portant *swara sancharas* (note transitions/movements) for the *rāgas* listed. Thus, MVRP listed by the musicians is not alongside the knowledge of the dataset considered, but independent of it. Hence, it is possible that the listed phrases may not appear in the specific compositions of the *rāgas* considered in the database.

Let $\mathcal{N}_{ref}^{(r)}$ represent the set of all MVRP for a *rāga* $r$. For RID using the MVRP, we assign a *rāga* label, $\hat{r}$ to test data as:

$$\hat{r} = \arg\max_{r \in \mathcal{R}} C(\mathbf{A}_{test} \mid \mathcal{N}_{ref}^{(r)}) \quad (8)$$

where $C(\mathbf{A}_{test} \mid \mathcal{N}_{ref}^{(r)})$ represents the count of occurrences of elements of $\mathcal{N}_{ref}^{(r)}$ in $\mathbf{A}_{test}$. However, unlike dictionary based approaches, $\mathbf{A}_{train}^{(r)}$ is not required to obtain $\mathcal{N}_{ref}$ set.

### 5.5 Performance metric

*5.5.1 RID as detection problem*

Viewing RID as a detection problem, precision and recall measures are used to gauge the performance of the algorithms. Let $TP^{(r)}$, $FP^{(r)}$ and $FN^{(r)}$ denotes the total number of true positive, false positive and false negatives of each *rāga* $r$, obtained by testing all the 4-folds; then, precision is $P^{(r)} = \frac{TP^{(r)}}{TP^{(r)}+FP^{(r)}}$ and recall is $R^{(r)} = \frac{TP^{(r)}}{TP^{(r)}+FN^{(r)}}$. The former is a measure of quality of correct detections and latter, a measure of quantity of correct detections.

*5.5.2 Perplexity of the models*

Perplexity (or $2^{cross\ entropy}$) measures goodness of model's distribution for given data (whose actual distribution is unknown). Maximizing probability implies minimizing perplexity. Lower perplexity implies a better model, and a perplexity of 1 on training data indicates over-fitting, as the model exactly represents the seen data and may not generalize to unseen test data. Drawing on the analogy between *rāgas* and languages, we quantify the performance of SMRNP techniques using perplexity, for both training and test data corpora. As true distribution of RNP ($p(\boldsymbol{x})$) is unknown, empirical distribution is used (Jurafsky and Martin, 2009). Perplexity of a model $\mathcal{D}^{(r)}$ w.r.t. data sequence $\mathbf{A}$ of $\mathbf{T}$ notes is:

$$\mathcal{P}^{(r)}(\mathbf{A}) = 2^{-\sum_{\boldsymbol{x} \in \mathbf{A}} p(\boldsymbol{x}) \log_2 \Pr(\boldsymbol{x}; \mathcal{D}^{(r)})} \approx 2^{-\frac{1}{\mathbf{T}} \sum_{t=1}^{\mathbf{T}} \log_2 \Pr(\boldsymbol{x}_t; \mathcal{D}^{(r)})} \quad (9)$$

## 6. Results and Analysis
### 6.1 RID: Performance comparison

Table 3 summarizes the performance accuracy of RID averaged across *rāgas* and 4-fold test set, using MVRP, TDMS, $N$-gram, $M$-skip-$N$-gram and $N$-multigrams with bigram used as a baseline performance. $N$-multigram model shows better recognition performance than $N$-gram model or $M$-skip-$N$-gram model; specifically, $N$-multigram model with $N = 6$ shows higher precision and recall compared to other models

| Models | $N$ | $P$ | $R$ |
|---|---|---|---|
| MVRP | - | 0.39 | 0.35 |
| TDMS | $\tau = 0.3\ s$ | **0.85** | **0.79** |
|  | $\tau = 0.01\ s$ (bigram) | 0.52 | 0.60 |
|  | $\tau = 1\ s$ | 0.69 | 0.73 |
|  | $\tau = 6\ s$ | 0.67 | 0.71 |
| $N$-gram | 2 | 0.83 | 0.77 |
|  | 3 | **0.90** | **0.85** |
|  | 4 | 0.68 | 0.73 |
|  | 5 | 0.67 | 0.75 |
|  | 6 | 0.65 | 0.75 |
|  | 7 | 0.69 | 0.68 |
|  | 8 | 0.73 | 0.64 |
| $M$-skip-$N$-gram | 2 ($M=1$) | 0.66 | 0.70 |
|  | 2 ($M=2$) | 0.66 | 0.75 |
|  | 3 ($M=1$) | **0.845** | **0.875** |
|  | 3 ($M=2$) | 0.82 | 0.81 |
| $N$-multigram | 5 | 0.92 | 0.83 |
|  | 6 | **0.93** | **0.85** |
|  | 7 | 0.77 | 0.79 |
|  | 8 | 0.87 | 0.79 |

**Table 3:** Avg. precision and recall rates for RID using different models

for the considered *rāgas*. We analyze the performance of each model for RID task.

*6.1.1 Analysis of MVRP for RID*

Some reasons accounting low $P$ and $R$ for RID using listed transcribed MVRP are:

- The MVRP may / may not occur in the dataset considered due to the approach adopted in listing MVRP i.e., the transcripts are based on recordings and adheres to it, while MVRP are listed without any such restraint. If $k$ of $\mathcal{N}_{ref}$ phrases occur in the corpus (exact match atleast once), an objective assessment of agreement of these MVRP w.r.t. the considered dataset is reported with average recall measure $Rec = k/|\mathcal{N}_{ref}|$ (in 4-fold validation setting) for both training and test corpus, for each *rāga* in first two columns of Table 4. $Rec$ on training corpus is higher than test corpus (due to training data containing more data). Some *rāgas* such as *Shankarabharana* show extremely low $Rec$ in training as well as test data set.
- Within the listed MVRP, it is observed that some of these elements are common across different *rāgas*. The renditions are different due to specific intonation, *gamakas* and a sample such scenario is depicted in Figure 4 for 2 *rāgas* *Shankarabharana* and *Thodi*. The MVRP listed may not be distinctive without the finer note position and the associated *gamakas*; these lead to confusion amongst *rāgas* for the test data.
- The presence/absence of augmented/reduced notes in either test corpus or MVRP poses a problem for exact pattern match.

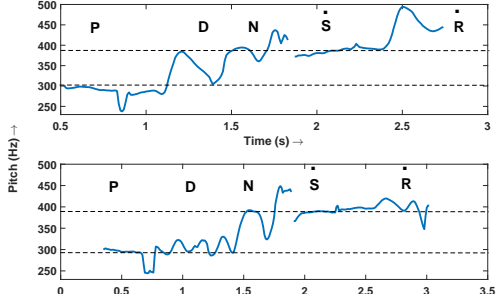

**Figure 4:** [Color Online] Pitch contour (in blue) of phrase '$PaDaNi\dot{S}a\dot{R}i$' rendered in *rāga Shankarabharana* and *Thodi* depicting the interplay of notes, intonation, *gamakas* and duration. The horizontal dashed lines indicate position of $Pa$ and $\dot{S}a$ with tonic at 200 Hz.

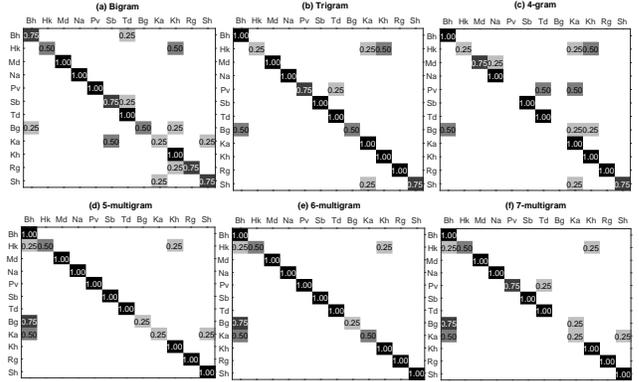

**Figure 5:** Confusion matrices for $N$-gram and $N$-multigram for varied $N$. *Rāga* abbreviations correspond to *rāga* names of Table 2.

### 6.1.2 RID using TDMS

The RID performance on equivalent audio data using TDMS technique is at 79%. The change in performance w.r.t. that reported in (Gulati et al., 2016) is due to changes in evaluation set (quality of recordings in the database, pitch extraction tool and length of recordings). TDMS algorithm essentially calculates bi-gram between the finely quantized melody contour and its $\tau$-delayed version. A post processed version of this $\tau$-delayed-bigram (cf. Section 7.2) is observed to result in an average of $\sim 25\%$ improvement in recall rate compared to that in the absence of post-processing ($\sim$52%) for $\tau = 0.3$ s. The performance of the algorithm w.r.t. delay parameter $\tau$ is tabulated. The $\tau$-delay-bigram model performs better than bi-gram model since a note would either have to stabilize or transitioned to the subsequent one after some delay. Similarly, increasing delay can affect delayed bi-gram statistics and hence reduce performance.

### 6.1.3 RID using N-gram models

The baseline performance of bigram is promising (and is close to that of TDMS at $\tau = 0.3$ s). Performance of trigram is better than that of bigram model. Increasing $N$ from trigram to 4-gram and further, shows a rapid degradation in performance. The average recall rate across *rāgas* decreases with increasing $N$, but precision does not show any consistent trend. For some *rāgas* such as *Begada* and *Panthuvarali*, detection of the said *rāgas* is not achieved for $N > 4$. This is due to unseen test sequences reducing likelihood of data w.r.t. model. This also accounts for trigram performing better than higher $N$-gram models. It can be safely said that trigram shows more promise than higher $N$-gram for RID task. We have also observed the confusion matrix changes markedly with varying $N$, resulting in varied precision rates. The same can be observed in Figure 5(a,b,c).

### 6.1.4 RID using M-skip-N-gram models

Skip-grams are $N$-grams formed from $M + N$ length contexts. In case of $M$-skip-2-gram models, increasing $M$ results in poorer performance, implying skipping more notes can alter statistics in an adverse way. 1-skip-trigram shows good $P$ and $R$ for RID; but for with $M \geq 2$, the distribution has reshaped in a degenerative way. Thus, on increasing analysis context ($M+N$), distribution of $N$-length sub-sequences is altered, but may not necessarily result in good RID performance. We hypothesize that in attempting to accommodate for improvisations, the statistics of adjacent notes is affected.

### 6.1.5 RID using N-multigram models

$P$ and $R$ of $N$-multigram models shows more promise than $N$-grams or $M$skip-$N$-gram models. A gradual performance degradation of $R$ as $N$ increases is observed. Within $N$-multigram models, RID accuracy is least for *rāga Begada* and *Kalyani*. Also the confusion matrices does not vary erratically with changing $N$, indicating that dictionaries across $N$ are consistent representations (cf. Fig 5(d,e,f)).

From confusion matrix of Figure 5(e), it can be observed *rāga Begada* and *Kalyani* are found confused with *rāga Bhairavi*, and *rāga Harikambhoji* is confused with *Bhairavi* and *Khamas*. Thus, though *Begada* and *Bhairavi* are rendered with different note positions (cf. Table 2), the dictionary representation obtained from 7-note data representation shows that the grammatical structure might be more similar than the rest (such as both *rāgas* having frequent usage of '$SaGaRiGa$' grammatical structure in forming a phrase).

From RID performance of SMRNP models, we can infer that RNP can characterize a *rāga*. Also, we infer that transcripts contain higher than 1st order Markov dependencies. While trigram and skip-trigram show competitive performance, $N$-multigram being a union of $\leq N$-grams gives a slight edge over the others and is less sensitive to choice of $N$. We look at perplexity measure in order to compare the models, and to see if it is indicative of RID performance.

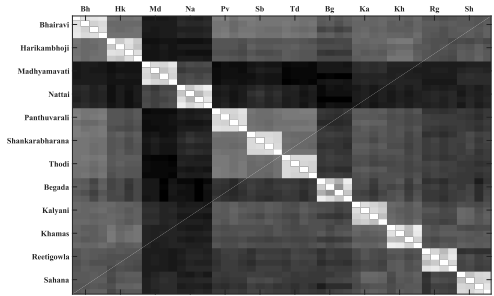

**Figure 6:** (a) Similarity matrix in inverted gray scale depicting the KL Divergence between 2 dictionaries obtained from $N$-multigram (N=6). Each *rāga* has 4 possible dictionaries (due to 4 fold CV).

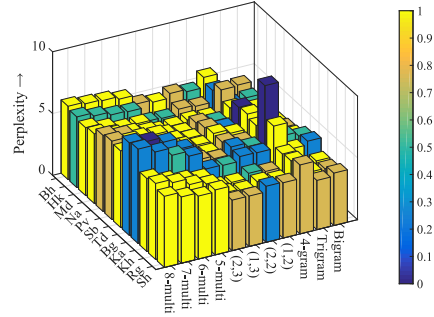

**Figure 7:** [Color online] *Rāga* -wise trends of $\mathcal{P}_{test}$ for SMRNP algorithms. The corresponding RID accuracy is indicated by the colorbar.

*6.1.6 Similarity measure for rāga dictionaries*

We analyze the distance between the $N$-multigram models to understand the similarities between the dictionaries. We define the similarity between $i^{th}$ and $j^{th}$ dictionary as the symmetric KL Divergence between the two i.e.,

$$\mathbb{S}(i,j) = \sum_{m=1}^{Y} \theta_m^{(i)} log \frac{\theta_m^{(i)}}{\theta_m^{(j)}} + \sum_{m=1}^{Y} \theta_m^{(j)} log \frac{\theta_m^{(j)}}{\theta_m^{(i)}} \qquad (10)$$

Figure 6 shows a similarity matrix obtained using symmetric KL divergence between all the folds of the dictionaries of all *rāgas*. Each dictionary of a *rāga* is compared with all other possible dictionaries. A block diagonal structure implies the dictionaries across the 4-fold training-sets are close to each other. Along the non-diagonal entries, we find that the dictionaries of *Begada* has least distance with those of *Bhairavi*, where as the *Begada* is not necessarily the closest to *Bhairavi*. This asymmetric nature of closeness can be observed by comparing any 2 rows of the KL divergence matrix.

## 6.2 Correlation of Perplexity with RID

Figure 7 depicts bar plot of *rāga*-wise perplexity, for different algorithms and with color-bar representing the RID recall rates for each *rāga*. While *rāga Begada* (Bg) and *Kalyani* (Ka) consistently show high perplexity trends, and their corresponding RID score is low for all SMRNP algorithms; however, when $\mathcal{P}_{test}$ is less, it does not imply better RID, as can be seen for *rāga Harikambhoji* (Hk). This implies that a SMRNP model may model unseen test data of a *rāga* well, however, it may not yet best represent a *rāga* w.r.t. other *rāga* models.

## 7. Contrasting the RID models

The SMRNP algorithms find repetitive temporal structures in note sequences within a local temporal window; $N$-gram considers consecutive $N$ notes, while skip-gram allows for $\leq M$ possible skips for considering $N$-length sequences and $N$-multigram considers all $\leq N$ sequences at initial iteration. We relate the algorithms considered. The relation between $N$-gram and $M$-skip-$N$-grams has been explicitly discussed in Section 3.2. Relation between other algorithms is reviewed here.

### 7.1 $N$-gram, $N$-multigram and Segment model

$\Pr(\mathbf{A})$ in $N$-gram model is obtained by applying chain rule (cf. (2)), where as in $N$-multigram model, it is obtained by a segmentation approach by formulation (cf. (3)). Consider (3) of $N$-multigram model; the note sequences within a metric cycle are a concatenation of varied length RNP. The independence of RNP sequences results in formulation as in (3). Clearly, it is a unigram of variable length RNP. If each RNP ($\boldsymbol{x}_o$) is a unit, a $N$-gram model applied to this sequence results in:

$$\Pr(A) = \Pr(\boldsymbol{x}_1, \boldsymbol{x}_2, \ldots, \boldsymbol{x}_{N-1}) \prod_{o=N}^{\omega} \Pr(\boldsymbol{x}_o | \boldsymbol{x}_{o-1} \ldots, \boldsymbol{x}_{o-N+1})$$

If $N$=2, then this results in first order Markov modeling of the RNP sequence. This is the basic segment model, also referred to as Hidden Semi-Markov models (HSMM), which are an extension to HMM (Hidden Markov Models), but with latent process being a semi-Markov chain. One way of introducing semi-Markov property is by explicitly modeling the duration of the states by a discrete distribution (Ostendorf et al., 1996); it is referred to as explicit duration HMM. Another approach to introduce semi-Markovian property has been the variable duration HMM, whose generalization is the segment model. Further, Hierarchical HMMs have been proposed as a generalization of segment models (Murphy, 2002a). Many more variations of HSMMs have been proposed (Yu, 2010; Johnson and Willsky, 2012). The basic segment model is the closest to multigram model. In segment models, each state of latent Markov chain emits a sequence of observations (while in HMM, each state emits a single observation vector). The difference between segment model and multigram models is that HSMM models have an additional Markov probability at the latent variable level, while multigram assumes independence of latent variables. Hence, multigram is a simpler, special case of segment model (Murphy, 2002b).

| Rāga | Rec on $A_{train}$ | Rec on $A_{test}$ | Rec on $\mathcal{D}_{N-gram}$ ||||  Rec on $\mathcal{D}_{N-multi}$ ||||
|---|---|---|---|---|---|---|---|---|---|---|
|  |  |  | N=5 | N=6 | N=7 | N=8 | N=5 | N=6 | N=7 | N=8 |
| Bh | 0.48 | 0.33 | 0.19 | 0.13 | 0 | 0 | 0.15 | 0.21 | 0.13 | 0.19 |
| Hk | 0.59 | 0.34 | 0.14 | 0.07 | 0.05 | 0.05 | 0.14 | 0 | 0.07 | 0.11 |
| Md | 0.42 | 0.23 | 0.22 | 0.03 | 0.03 | 0 | 0.08 | 0.03 | 0.03 | 0.03 |
| Na | 0.43 | 0.27 | 0.18 | 0 | 0.07 | 0 | 0.18 | 0.05 | 0.07 | 0.07 |
| Pv | 0.48 | 0.27 | 0.15 | 0.12 | 0.12 | 0 | 0.17 | 0.05 | 0.17 | 0.12 |
| Sb | 0.30 | 0.22 | 0.11 | 0 | 0.03 | 0 | 0.11 | 0.05 | 0.04 | 0.07 |
| Td | 0.32 | 0.2 | 0.15 | 0.12 | 0 | 0 | 0.11 | 0.07 | 0.03 | 0 |
| Bg | 0.43 | 0.21 | 0.23 | 0.05 | 0 | 0 | 0.18 | 0.18 | 0.11 | 0 |
| Ka | 0.52 | 0.26 | 0.20 | 0.10 | 0.05 | 0 | 0.15 | 0.05 | 0.07 | 0.02 |
| Kh | 0.38 | 0.18 | 0.1 | 0.05 | 0 | 0 | 0.05 | 0.05 | 0 | 0.05 |
| Rg | 0.52 | 0.37 | 0.33 | 0.05 | 0 | 0 | 0.33 | 0.05 | 0.05 | 0.05 |
| Sh | 0.38 | 0.22 | 0.27 | 0 | 0 | 0 | 0.18 | 0 | 0 | 0 |
| Avg. | **0.44** | **0.26** | **0.19** | **0.06** | **0.03** | **0.004** | **0.15** | **0.07** | **0.06** | **0.06** |

**Table 4:** The average recall rate of MVRP in training set, testing set and with $\mathcal{D}_{N-gram}$ and $\mathcal{D}_{N-multigram}$. In the latter two cases, $N$ is varied from $5$ to $8$.

### 7.2 TDMS technique and bi-gram

To relate TDMS algorithm with the bigram, we interpret the final 2 steps of TDMS algorithm for extracting features as follows: each element of the surface indicates the count of number of joint occurences of sequences (quantized levels) $(i, j)$ for all possible indices, but these quantized melodic points are not adjacent but separated by a time delay $\tau$ - this can be seen as $\tau$ delayed bigram (Huang et al., 1992) containing joint probabilities. The post processing is equivalent to a smoothing operation applied to $N$-grams to account for unseen sequences. Hence, we can refer to the algorithm is $\tau$-delayed-bi-gram.

### 7.3 Correlation of dictionaries of SMRNP with MVRP

In order to measure correlation of dictionary RNP (obtained from training data) with that of MVRP [8], we use standard precision, recall metrics (Meredith, 2016) used in retrieval applications to quantify the extent the dictionaries represent a set of MVRP of each *rāga*. . If $k$ of $\mathcal{N}_{ref}$ MVRP are detected in a *rāga* dictionary of size $|\mathcal{D}_{model}|$, then precision, $Prc \triangleq k/|\mathcal{D}^{(r)}|$ and recall, $Rec = k/|\mathcal{N}_{ref}^{(r)}|$. Owing to large size of dictionaries in either $N$-gram (of order of $>10,000$) or in $N$-multigram (of order of $3000-4000$), the precision measure is bound to be low and hence we report only recall for each *rāga*, averaged over the dictionaries obtained in the 4-fold validation setting. The average recall rate for all *rāgas* and for varied values of $N$ is summarized in Table 4.

$Rec$ in $\mathcal{D}_{N-gram}$ is non-zero if a listed $N$ length

[8] cf. https://sites.google.com/site/carnaticmvrp/

MVRP is present in training set. Trivially, the sum of the columns of $Rec$ on $\mathcal{D}_{N-gram}$ must be less than or equal to that of the training set; $\mathcal{D}_{N-gram}$ contains all $N$ length RNP, each assigned a non-zero probability either due to presence in training data or due to smoothing. Amongst, $N$-grams, 5-gram shows highest non-zero recall rates, indicating many *rāgas* have elements in MVRP of length 5. The average $Rec$ across *rāgas* drops drastically for $N=6$-gram and beyond. Average $Rec$ using $\mathcal{D}_{N-multi}$ degrades gradually with increasing $N$. For some *rāgas*, $Rec$ is sometimes lesser than that of $N$-gram as RNP of $\mathcal{D}_{N-multi}$ must necessarily occur atleast once in the dataset (and there is no smoothing). The segmentation step in $N$-multigram models weigh higher length sequences ($N=6,7,8$) over lower length ones ($N <= 5$). This also contributes to lower recall rates. However, in some *rāgas*, $Rec$ on $\mathcal{D}_{N-multi}$ is higher than $N$-gram as $\mathcal{D}_{N-multi}$ includes RNP of length $<= N$. The same also reflects as lack of monotonic behavior in $Rec$ on $\mathcal{D}_{N-multi}$ with increasing $N$. Due to validity of both possibilities, we can infer that there is not much overlap between discovered RNP and MVRP (the latter is more dependent on musicological knowledge than the former).

### 7.4 Analysis of Dictionary

In $N$-multigram models, effective size of dictionaries can be different for various *rāgas*, and in each validation fold of a *rāga*; whereas the dictionary size of $N$-gram and $M$-skip-$N$-gram ($M$ plays a role in constructing more $N$-grams) is same for all *rāgas* and is equal to $\mathbb{V}^N$. For example, size of $5-$gram dictionary will be $7^5 = 16807$ and exponential growth is expected with increasing $N$. The dictionaries of $N$-gram (and skip-grams) are sparsely populated, but all unencountered entries are assigned non-zero (and miniscule) values as a result of Good-Turing smoothing. In contrast, dictionary size in $N$-multigram is obtained from optimal segmentation. A plot of average dictionary sizes (from the dictionaries of 4-fold training set) for each *rāga* is shown in Figure (8). As expected, dictionary size of $N$-multigram for any *rāga* increases with increasing $N$. Comparing with the $N$-gram dictionary size, the number of entries in $N$-multigram does not exceed $5000$ even for $N = 8$ parameter. Thereupon, $N$-multigram model can be said to be a more compact representation than $N$-gram (or skip-gram).

A sample clip of dictionaries, $\mathcal{D}_{N-gram}$ and $\mathcal{D}_{N-multi}$ for *Begada rāga* is shown in Table 5 for $N = 6$. Some of the successive entries of $\mathcal{D}_{N-gram}$ (Ex: Entries 4,5,7 or 14-16) are part of a longer sequence. Many of the RNP have same count resulting in them having same probability values.

$\mathcal{D}_{N-multi}$ contains RNP of length ranging from $1$ to $N$. The probability values in $\mathcal{D}_{N-multi}$ is higher than that of $\mathcal{D}_{N-gram}$ for the same RNP; for example, the RNP $MaMaGaRiRiPa$ is found by both the models, but $\mathcal{D}_{N-gram}$ assigns a probability of $0.004$ (entry 15), while the same RNP is assigned a higher value, 0.017, in $N$-multigram (entry 4). Thus, inspite of presence of some common RNP in $\mathcal{D}_{N-multi}$ and $\mathcal{D}_{N-gram}$, probabilities of RNP are different. In $N$-gram models, the count is normalized by the total number of valid $N$ length sequences, whereas in $N$-multigram, the count is normalized by $\Omega(\mathfrak{b}^{(*,r)})$. The final number of segments determined by the $\mathcal{D}_{N-multi}^{(r)}$ - $\Omega(\mathfrak{b}^{(*,r)})$, is unknown initially and corpus dependent, while $Y_{N-gram}^{(r)} = |\mathbb{V}^N|, \forall r$ is corpus independent. This accounts for differences in the estimated distributions. On analyzing $\mathcal{D}_{N-multi}^{(r)}$, w.r.t. the training data, some RNP are observed to be truncated due to maximum RNP length limited by $N$. Increasing $N$ will increase computational complexity on one end and also leads to forcefully looking for longer sequences, resulting in over-training.

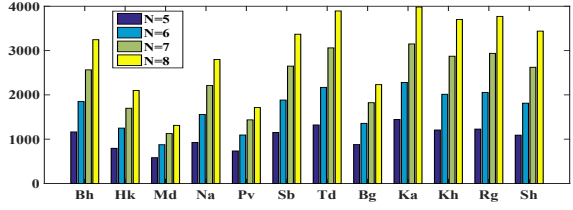

**Figure 8:** [Color Online] Average dictionary size using $N$-multigram for various *rāgas* for $N$ varying from 5 to 8. Contrastingly $N$-gram ($M$-skip-$N$-gram) dictionaries are of size 16807, 117649, 823543 and 5764801 for $N$ varying from 5 to 8, but sparsely populated.

| Sl.no | RNP of $\mathcal{D}_{N-gram}$ | Pr($s$) | RNP of $\mathcal{D}_{N-multi}$ | Pr($s$) |
|---|---|---|---|---|
| 1 | Pa Pa Pa Pa Pa Pa | 0.022 | Pa Pa Pa Pa Pa Pa | 0.031 |
| 2 | Sa Sa Sa Sa Sa Sa | 0.010 | Sa Sa Sa Sa Sa Sa | 0.020 |
| 3 | Pa Ma Pa Da Pa Sa | 0.004 | Ma Ma Ma | 0.017 |
| 4 | Ni Ni Ni Da Pa Pa | 0.004 | Ma Ma Ga Ri Ri Pa | 0.017 |
| 5 | Pa Pa Ma Pa Da Pa | 0.004 | Ma Pa Da Pa Sa Ni | 0.017 |
| 6 | Ni Ni Ni Ni Da Pa | 0.004 | Sa Ni Da Pa Sa | 0.011 |
| 7 | Ni Ni Ni Ni Ni Da | 0.004 | Da Pa Ma Pa Pa | 0.011 |
| 8 | Ri Sa Sa Sa Sa Sa | 0.004 | Pa Pa Pa Pa Pa | 0.011 |
| 9 | Ni Sa Ri Sa Ri Sa | 0.004 | Ma Ni Da Pa Pa Pa | 0.011 |
| 10 | Ni Sa Ni Sa Ri Sa | 0.004 | Ni | 0.008 |
| 11 | Sa Ni Sa Ri Sa Ri | 0.004 | Da Pa | 0.008 |
| 12 | Ma Ma Ma Ga Ri Ri | 0.004 | Sa Ri Sa | 0.008 |
| 13 | Ma Ma Ma Ma Ga Ri | 0.004 | Ni Sa Ni | 0.008 |
| 14 | Ma Ma Ma Ma Ma Ga | 0.004 | Ni Sa Ni Sa | 0.008 |
| 15 | Ma Ga Ga Ri Ri Pa | 0.004 | Ni Ni Da Pa | 0.008 |
| 16 | Da Pa Pa Pa Pa Pa | 0.004 | Pa Da Pa Ri Sa | 0.008 |
| 17 | Sa Ri Sa Ri Sa Ni | 0.004 | Ni Da Pa Ni Ri Sa | 0.008 |
| 18 | Ma Pa Da Pa Sa Ni | 0.004 | Ni Ni Ni Ni Da Pa | 0.008 |
| 19 | Pa Sa Sa Sa Sa Sa | 0.003 | Ri Sa Ri Sa Ni Da | 0.008 |
| 20 | Pa Ri Sa Sa Sa Sa | 0.003 | Ni Ni Ni Ni Ni Da | 0.008 |

**Table 5:** [Color Online] Dictionaries of *Begada rāga* for $N$=6 sorted in decreasing order of probabilities. Some MVRP are highlighted in blue color.

The 4-fold training results in 4 (different) dictionaries, for a given *rāga*. Owing to data dependency, only a few RNP of $\mathcal{D}_{N-multi}^{(r)}$ across the 4 dictionaries will be common. For example, the number of common RNP in all the 4-fold dictionaries of *rāga Begada* are much lesser ($< 100$) compared to the size of dictionary depicted in Figure 8.

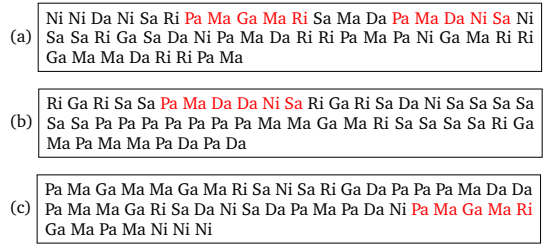

**Figure 9:** [Color Online] Music notations generated for *Sahana rāga* using (a) $\mathcal{D}_{multi}$ with $N = 6$ (b) $\mathcal{D}_{N-gram}$ with $N = 3$ (c) $\mathcal{D}_{skip}$ with $(M, N) = (3, 1)$. MVRP that can be manually identified from the generated sequences are highlighted in red.

A few MVRP present in $\mathcal{D}^{(r)}$ are identified (highlighted using blue font in Table 5. A detailed analysis of correlation between dictionaries and MVRP is presented in Section 7.3.

### 7.5 Generating sequences

As the SMRNP considered are generative models, we generate music notations. A sample sequence generated by the models for *Sahana rāga* is shown in Figure 9.

## 8. Discussion and Conclusions

Inspite of low match with MVRP, RID results demonstrate that RNP obtained from using just 7 notes of skeletal transcripts are a better feature representation than MVRP itself and can identify a *rāga*. The RID results of SMRNP algorithms are an indicator that skeletal note transcripts contain considerable amount of *rāga* information. Also, the sequencing of notes itself can aid in *rāga* identification task. Higher order($> 1$) Markov dependencies are present in the data as can be seen from performance of trigram, skip-gram and $N$-multigram techniques. We hypothesize that inclusion of finer note interval (such as $Ri_1; Ri_2; Ri_3$) will automatically boost RID accuracy. Such grouping of finer note sequences complemented with knowledge of *gamakas* can result in better *rāga* models. This would be akin to the effect of adding grammar models to acoustic models in speech recognition.

The melody contour based TDMS algorithm essentially utilizes very fine note positions, while smoothing out effects of durations in obtaining transition matrix from the melody signal (Gulati et al., 2016). It does not utilize sequential information. Melodic contours are a florid representation of prescriptive notation replete with note intonations, durational informations and *gamakas*. Yet, this large amount of information can pose difficulties in computationally representing/identifying a *rāga*. This is reflected in the recall rates using TDMS technique. There is potential promise in using $N$-grams/$N$-multigrams with $N > 2$ in TDMS algorithm. This is not trivial as it will involve more than 2-dimensional surfaces. While we have explored the advantages of utilizing (prescriptive) music notations for characterizing a *rāga*, the problem of mapping melodic contours to symbolic notation and vice versa is still in its nascent stage [9]. It essential for a detailed understanding and representation of the music form.

In this work, the notes across metric cycles are assumed to be independent. While MVRP can exist across metric cycles, the assumption permits us to leverage the computational advantage in the segmentation stage of the multigram model. The results (with the independence assumption) show the importance of considering long variable length sub-sequences for characterizing a *rāga*.

The effects of fine durational aspects of notation (such as reduced and augmented note durations) are not considered in this work. Also, many of the compositions contain morphed MVRP with some notes augmented or reduced in duration (the notes that allow duration reduction and/or augmentation are based on grammar of a *rāga*), so as to fit well in the metric cycle. Future work must incorporate these to result in a better dictionary representation and its correlate with MVRP. Segmenting such patterns and incorporating beat information into its unique representation will result in sophisticated models for the music form. We have examined and evaluated the importance of temporal structure in Carnatic music notations although it is considered only as a skeleton of a rendition. The study shows that it is possible to automatically identify a *rāga* based on recurrent note patterns. Although, few MVRP are found in the stochastic dictionaries, these alone do not suffice to characterize *rāga* for RID task. Experiments show that $N$-multigram model provides advantages over $N$-gram models as well as skip grams in capturing RNP of varied lengths; in terms of perplexity, they are a good fit to test data and result in better performance on RID task. Also, we have shown that it is not necessarily beneficial to increase the length of RNP ($N$). Owing to the performance, $N$-multigram dictionary can be viewed as an viable symbolic feature representation of a *rāga*.

Grammatical structures are present internally in the prescriptive notation and seem to play a stronger role than the fine pitch values themselves as can be contrasted in the performance of $N$-multigram with that of TDMS. The stochastic language models explored in this work are generic but analyzed specific to Carnatic music form in this work; Carnatic music contains more ornamentations than Hindustani music form. As we have been able to capture *rāga* structure with use of only 7 notes of the prescriptive notations, we hypothesize the analysis will hold for other music forms such as Hindustani and Turkish classical music where phrases form a major basis for melody description. While $N$-gram models are employed in analysis of folk music, Western music and segmentation algorithms for structural segmentations (Sargent et al., 2017; Levy and Sandler, 2006), we believe the usage of segment model in modeling grammar and in evaluating melodic similarity in other forms of music such as Jazz will be analogous.

Summarizing, the major contributions of this work are:

1. Starting from the octave-folded notations of *rāgas* in Carnatic music, using only 7 basic notes and no *gamaka* information, we show that it is possible to identify *rāgas* based only on the statistics of sub-sequences. This demonstrates the validity of RNP as a representatives of *rāgas*.
2. For the set of *rāga* renditions containing both audio recordings and notations, we show that *rāga* identification using the notations can outperform the state of the art audio-only based technique for *rāga* identification. This signifies result is that we can see that much of *rāga* structures are present as discrete elements while expressions are present as continuous variabilities.
3. Experimental evidence to show that the proposed approach (based on repetitive patterns) outperforms identifying *rāgas* that based on musicologically valid reference phrases, as agreed upon by three practicing musicians. This shows that repetitive patterns are more powerful than characteristic patterns for computational analysis.

## A. Ratio of notes within an octave

The tonic frequency in Indian art music is the reference base frequency. The rest of the notes are fixed with respect to this anchor. The ratios between the notes within an octave are maintained irrespective of the starting tonic pitch frequency (Krishnaswamy, 2003). Table (6) tabulates the ratios and frequencies of notes w.r.t. a fixed tonic $f_0 = 146.83\ Hz$ (corresponding to note in Western music, $D_3$). A melodic notation/transcription cannot have the following pairs as they map to the same frequency: $(Ri_2, Ga_1)$, $(Ri_3, Ga_2)$, $(Da_2, Ni_1)$ or $(Da_3, Ni_2)$.

## Acknowledgments

We thank musicians Vid. Deepak Paramashivan, Vid. Raman Shankaran (violinist) and Vid. Deepika Janakiraman (vocalist) for helping create MVRP list. We also thank authors of (Gulati et al., 2016) for providing the original implementation of their algorithm.

---

[9] The mapping of pitch contours to its *swara* notation can be (i) manual - involves a trained musician transcribing the audio clip to its corresponding notes. (ii) automated - note sequences can be obtained by the process of quantizing pitch contours of the lead melody extracted from the audio signal of a Carnatic music rendition (Ranjani et al., 2011, 2017) (Koduri et al., 2014). The challenges in automatic transcription is detailed in (Krishnaswamy, 2003; Koduri, 2016).

| | Notes | $f_0$ in Hz (ratio with tonic) |
|---|---|---|
| 1. | $Sa$ | 146.83 (1) |
| 2. | $Ri_1$ | 156.66 (1.067) |
| 3. | $Ri_2$ or $Ga_1$ | 165.18 (1.125) |
| 4. | $Ri_3$ or $Ga_2$ | 176.19 (1.20) |
| 5. | $Ga_3$ | 183.53 (1.25) |
| 6. | $Ma_1$ | 195.28 (1.33) |
| 7. | $Ma_2$ | 207.9 (1.416) |
| 8. | $Pa$ | 220.24 (1.5) |
| 9. | $Da_1$ | 234.92 (1.6) |
| 10. | $Da_2$ or $Ni_1$ | 244.76 (1.667) |
| 11. | $Da_3$ or $Ni_2$ | 264.29 (1.8) |
| 12. | $Ni_3$ | 275.30 (1.875) |
| 13. | $\dot{S}$ | 293.66 (2) |

**Table 6:** Relative pitch positions, frequencies and ratio of notes within an octave.